\documentclass[conference]{IEEEtran}
\IEEEoverridecommandlockouts
\usepackage[utf8]{inputenc} 
\usepackage[T1]{fontenc}
\usepackage{url}
\usepackage{ifthen}
\usepackage{cite}
\usepackage[cmex10]{amsmath} % Use the [cmex10] option to ensure complicance
\usepackage{textcase}
\usepackage[tablename=Table]{caption}
\usepackage{blindtext}
\usepackage{floatrow}
\usepackage{soul}
\usepackage{changes}
\usepackage{gensymb}
\usepackage{cite}
\usepackage{amsmath,amssymb,amsfonts}
\usepackage{textcomp}
\usepackage{graphicx}
\usepackage{amssymb}
\usepackage{amsfonts}
\usepackage{amsmath}
\usepackage{epsfig}
\usepackage{color}
\usepackage{fancybox}
\usepackage{textcomp}
\usepackage{multirow}
\usepackage{setspa ce}
\usepackage{psfrag}
\usepackage{booktabs}
\usepackage{float}
\usepackage{algorithm}
\usepackage{algpseudocode}
\usepackage{mathtools, nccmath, bigints, amsfonts}

\usepackage{caption}
\usepackage{subcaption}
\usepackage{mathrsfs}								% 支持花体（大写）

\usepackage{placeins}

\newfloatcommand{capbtabbox}{table}[][0.4\textwidth]

\newtheorem{Remark}{Remark}
    \def\Complex{{\rm\rule[.23ex]{.03em}{1.1ex}\kern-.3em{C}}}

    \newcommand{\be}{\begin{equation}} \newcommand{\ee}{\end{equation}}
    \newcommand{\bea}{\begin{eqnarray}} \newcommand{\eea}{\end{eqnarray}}
    \newcommand{\benum}{\begin{enumerate}} \newcommand{\eenum}{\end{enumerate}}

        \newcommand{\qh}{{\bf h}}

        \newcommand{\qr}{{\bf r}}
        \newcommand{\qs}{{\bf s}}

        \newcommand{\qw}{{\bf w}}
        \newcommand{\qx}{{\bf x}}
        \newcommand{\qy}{{\bf y}}

        \newcommand{\qF}{{\bf F}}
        \newcommand{\qG}{{\bf G}}
        \newcommand{\qH}{{\bf H}}
        \newcommand{\qI}{{\bf I}}

        \newcommand{\qR}{{\bf R}}

        \newcommand{\qX}{{\bf X}}
        \newcommand{\qY}{{\bf Y}}

        \newcommand{\qDelta}{{\boldsymbol \Delta}}

        \newcommand{\qSigma}{{\boldsymbol \Sigma}}

        \newcommand{\qmu}{{\boldsymbol \mu}}

        \newcommand{\Ex}{{\sf E}}

\def\BibTeX{{\rm B\kern-.05em{\sc i\kern-.025em b}\kern-.08em
    T\kern-.1667em\lower.7ex\hbox{E}\kern-.125emX}}
\begin{document}

\title{Bayesian-based Symbol Detector for Orthogonal Time Frequency Space Modulation Systems} 

 \author{%
   \IEEEauthorblockN{
   					Xinwei Qu\IEEEauthorrefmark{1},
   					Alva Kosasih\IEEEauthorrefmark{1},
                    % Vera Miloslavskaya\IEEEauthorrefmark{1},
                     Wibowo Hardjawana\IEEEauthorrefmark{1}, 
                    % Victor Andrean\IEEEauthorrefmark{2},
                    Vincent Onasis\IEEEauthorrefmark{1},
                    and Branka Vucetic\IEEEauthorrefmark{1}\\
                    }
   \IEEEauthorblockA{\IEEEauthorrefmark{1}%
                     Centre of Excellence in Telecommunications, University of Sydney, Sydney, Australia.  \\
                    \{alva.kosasih,wibowo.hardjawana,branka.vucetic\}@sydney.edu.au,\\
                     xiqu4217@uni.sydney.edu.au,  and  vona0880@uni.sydney.edu.au.   }
                    
   %\IEEEauthorblockA{\IEEEauthorrefmark{2}%
                    % Mobilizing Information Technology Lab., National Taiwan University of Science and Technology, Taipei, Taiwan. }
 }

\maketitle

%%%%%%
%% Abstract: 
%% If your paper is eligible for the student paper award, please add
%% the comment "THIS PAPER IS ELIGIBLE FOR THE STUDENT PAPER
%% AWARD." as a first line in the abstract. 
%% For the final version of the accepted paper, please do not forget
%% to remove this comment!
%%
\begin{abstract}

Recently, the orthogonal time frequency space (OTFS) modulation is proposed for 6G wireless system to deal with high Doppler spread. The high Doppler spread  happens when the transmitted signal is reflected towards the receiver by fast moving objects (e.g. high speed cars), which causes inter-carrier interference (ICI). 
%To date, existing 4/5G OFDM transmission technology can not cancel this type of inter-carrier interference.  
%To overcome such  ICI, the orthogonal time frequency space (OTFS) modulation  has been proposed, recently.
Recent state-of-the-art OTFS detectors fail to achieve an acceptable bit-error-rate (BER) performance as the number of mobile reflectors increases which in turn, results in high inter-carrier-interference (ICI).
% due to the presence of high interference or suffer a high latency due to the need to use more cyclic prefix. The latter is not preferred, as data traffic in mobile cellular networks are latency sensitive.
In this paper, we propose a novel detector for OTFS systems, referred to as the Bayesian based parallel interference and decision statistics combining (B-PIC-DSC) OTFS detector that can achieve a high BER performance, under high ICI environments. The B-PIC-DSC OTFS detector employs the PIC and DSC schemes to iteratively cancel the interference, and the Bayesian concept to take the probability measure into the consideration when refining the transmitted symbols. Our simulation results show that in contrast to the state-of-the-art OTFS detectors, the proposed detector is able to achieve a BER of less than $10^{-5}$, when SNR is over $14$ dB, under high ICI environments.
\end{abstract}


%% The paper must be self-contained. However, if you are referring to
%% a full version for checking certain proofs, please provide the
%% publically accessible location below.  If the paper is completely
%% self-contained, you can remove the following line from your
%% submission.

\section{Introduction}

\begin{figure}
    \centering
    \includegraphics[width=0.85\textwidth]{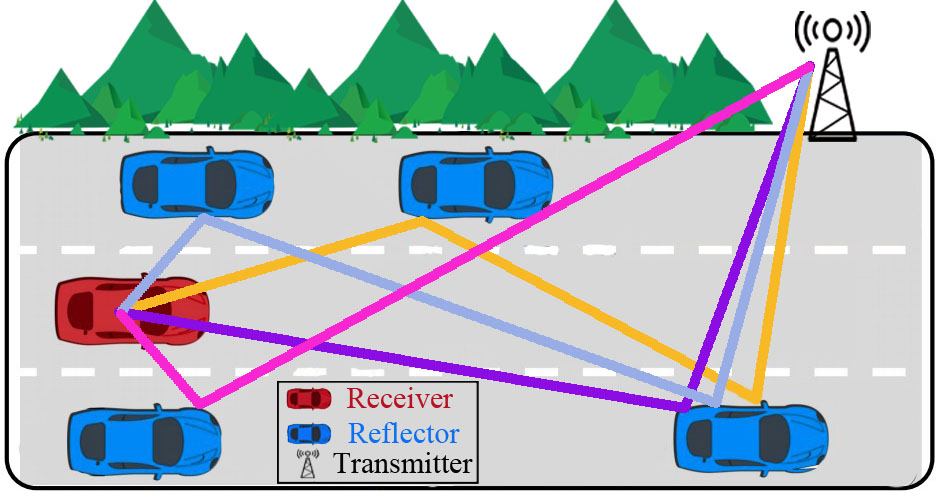}
    \caption{High mobility reflectors}
    \label{fig:intro-high_mob_ref}
\end{figure}

The sixth-generation ($6$G) wireless system will support advance mobile network applications (e.g. unmanned aerial vehicles (UAV), and autonomous cars) \cite{hashimoto2020channel, wang20206g}, which come with stringent error-rate requirement  under high Doppler spread environments. For example, in the autonomous car application where there exists high mobility reflectors (i.e. the other moving vehicles around the vicinity of a specific car receiver), high Doppler spread is observed. This is shown in Fig. \ref{fig:intro-high_mob_ref}. These reflectors cause an inter-carrier interference (ICI) which significantly degrades the performance of the current orthogonal frequency division multiplexing (OFDM) systems\cite{jiang2010channel}. The orthogonal time frequency space (OTFS) modulation has been proposed in\cite{hadani2017orthogonal} to tackle this issue by multiplexing the transmitted symbols in the delay-Doppler (DD) domain and then spreading them out in the time-frequency (TF) domain. 
%To gain full diversity in the TF domain, an OTFS symbol detector should be able to achieve high detection performance with a reasonable complexity\cite{hadani2017orthogonal}. 

The performance of classical low complexity detectors, i.e., zero forcing (ZF) and minimum-mean-square-error (MMSE) detectors, in the OTFS system have been investigated in \cite{singh2020low}. 
%They use a  matrix inverse operation to detect the transmitted symbols from the received signal. 
 Despite low complexity, the classical detectors suffer a significant performance degradation, compared to the optimal maximum likelihood (ML) detector\cite{murali2018otfs}. To address this issue, iterative receiver algorithms have been employed for the OTFS detectors based on the message passing (MP) \cite{raviteja2018interference},  approximate message passing (AMP), and unitary transformation AMP (UTAMP)   \cite{yuan2020iterative}.  The MP based detectors, in the OTFS system, use a message passing iterative algorithm to compute the posterior probability function of the transmitted symbols, where the messages represent Gaussian approximations of the ICI.
% The MP detector can achieve high detection performance as it was developed based on  the belief propagation detector which can be regarded as the optimal detector in a tree graphical model\cite{goldberger2011mimo}.
% J. Yedidia,W. Freeman, and Y. Weiss, “Understanding belief propagation %and its generalizations,” Exploring artificial intelligence in the new
%millennium, vol. 8, pp. 236–239, 2003.
Unfortunately, the BER performance of the MP based detectors degrades due to the increase in ICI as the number of moving reflectors rises. 
%In addition, the UTAMP OTFS detector requires diagonalizable channel matrix \cite{yuan2020iterative} which limits its  implementation. 
Recently, we proposed an iterative Bayesian based detection algorithm, referred to as the Bayesian parallel interference and decision statistic combining (B-PIC-DSC). The algorithm iteratively estimates the mean and variance of the transmitted symbols and uses them to subtract interference from the received signal. The proposed algorithm is shown to achieve a significant bit-error-rate (BER) performance gain, as compared to the state-of-the-arts, in the massive multiple-input-multiple-output scenario\cite{kosasih2020linear}. 

%However, the performance of the MP based detectors degrade significantly under high interference environments\cite{caltagirone2014convergence}, such as in high ICI environments. Moreover, the UTAMP OTFS detector requires diagonalizable channel matrix which limits  its  implementation.
%Aiming to solve this problem, an advance approximate message passing  (AMP) detector, called the unitary transformation AMP  detector has been proposed for OTFS system, referred to as  the UTAMP OTFS detector\cite{yuan2020iterative}. The main idea of the UTAMP OTFS detector is to diagonalize the channel matrix in the DD domain using the 2D fast Fourier transform (FFT) algorithm which causes a significant latency increase. 
%Nevertheless, the diagonalization process requires inserting more cyclic prefixes compared to the efficient OTFS system model in \cite{raviteja2018practical}. This leads to a significant latency increase  which is not preferable for mobile cellular networks. 

{In this paper, we propose a novel iterative OTFS detector. The detector is based on the B-PIC-DSC detection algorithm \cite{kosasih2020linear}, referred to as the B-PIC-DSC OTFS detector. 
The B-PIC-DSC OTFS detector first applies the PIC scheme to remove the ICI from the received signals using the symbol estimates from the previous iteration. A factorizable Gaussian posterior belief of the transmitted symbols is then constructed based on the PIC outputs. The posterior belief of the transmitted symbols is then used to infer the soft symbol estimates. The DSC concept is then utilized to yield the final symbol estimates by weighting the soft symbol estimates in the previous and current iterations.  The process is then repeated iteratively until there is no significant performance improvement in the soft symbol estimates. The DSC outputs, from the last iteration, are then used to refine the transmitted symbols. 
%Note that we follow the OTFS model, considered in \cite{raviteja2018practical}, where the cyclic prefix (CP) is only inserted in the beginning of the OTFS frame,  regarded as the most efficient OTFS model in terms of latency and spectral efficiency.
%The main contributions are summarized as follows:
%We aim to significantly improve the performance of the current state-of-the-art MMSE based detectors with the same computational complexity order.
The main contribution of this paper is the development of the B-PIC-DSC detector for an OTFS system  that can achieve a high BER performance, in the presence of a strong ICI due to a large number of moving reflectors. The simulation results demonstrate that our proposed detector is able to achieve a BER of $10^{-5}$, in the presence of a large number of high mobile reflectors. This is in contrast to other OTFS detectors that fail to work in such an environment.

{\bf Notations}: $a$, $\textbf{a}$ and $\textbf{A}$ denote scalar, vector, and matrix respectively. $\mathbb{C}^{M\times N}$ denotes the set of $M\times N$ dimensional complex matrices. We use $\qI_N$, $\textbf{F}_N$, and $\textbf{F}_N^{\qH}$ to represent an $N$-dimensional identity matrix, $N$-points discrete Fourier Transform matrix, and $N$-points inverse discrete Fourier transform matrix. $(\cdot)^T$, $(\cdot)^{\qH}$, $(\cdot)^*$, and $[\cdot]_M$ represent the transpose, Hermitian,  conjugate, and mod-$M$ operations. We define $\textbf{a} = {\sf vec}(\textbf{A})$ as the column-wise vectorization of matrix $\textbf{A}$.
%Notation $diag(\textbf{a})$ denotes a diagonal matrix whose diagonal is the vector $\textbf{a}$, $diag(\qA)$ denotes a vector that is the diagonal of the matrix $\textbf{A}$ and $diag_0(\qA)$ denotes the operation that forces the diagonal of matrix $\qA$ to zero. 
The Kronecker product is denoted as $\otimes$. 
% , and $\qa \cdot \qb$, $\qa ./ \qb$ and $|\cdot|^2$ denote the element-wise production production, division and magnitude squared operation respectively. 
% We use $\bf 0$ and $\bf 1$ to represent the adequately long vectors full of 0 and 1 respectively. The superscript $(\cdot)^{(t)}$ denotes the $t$-th iteration and 
The Euclidean distance of vector $\qx$  is denoted as $\|\qx\|$. We use $\mathcal{N}(\qx: \qmu, \qSigma)$ to express the multivariate Gaussian distribution of a vector $\qx$ where $\qmu$ is the mean and $\qSigma$ is the covariance matrix.

%% <Existing Words>
% OTFS OFDM, [.]_M
\section{System Model}

\begin{figure}
\centering
    \begin{subfigure}[b]{0.6\textwidth}
         \centering
         \includegraphics[width=\textwidth]{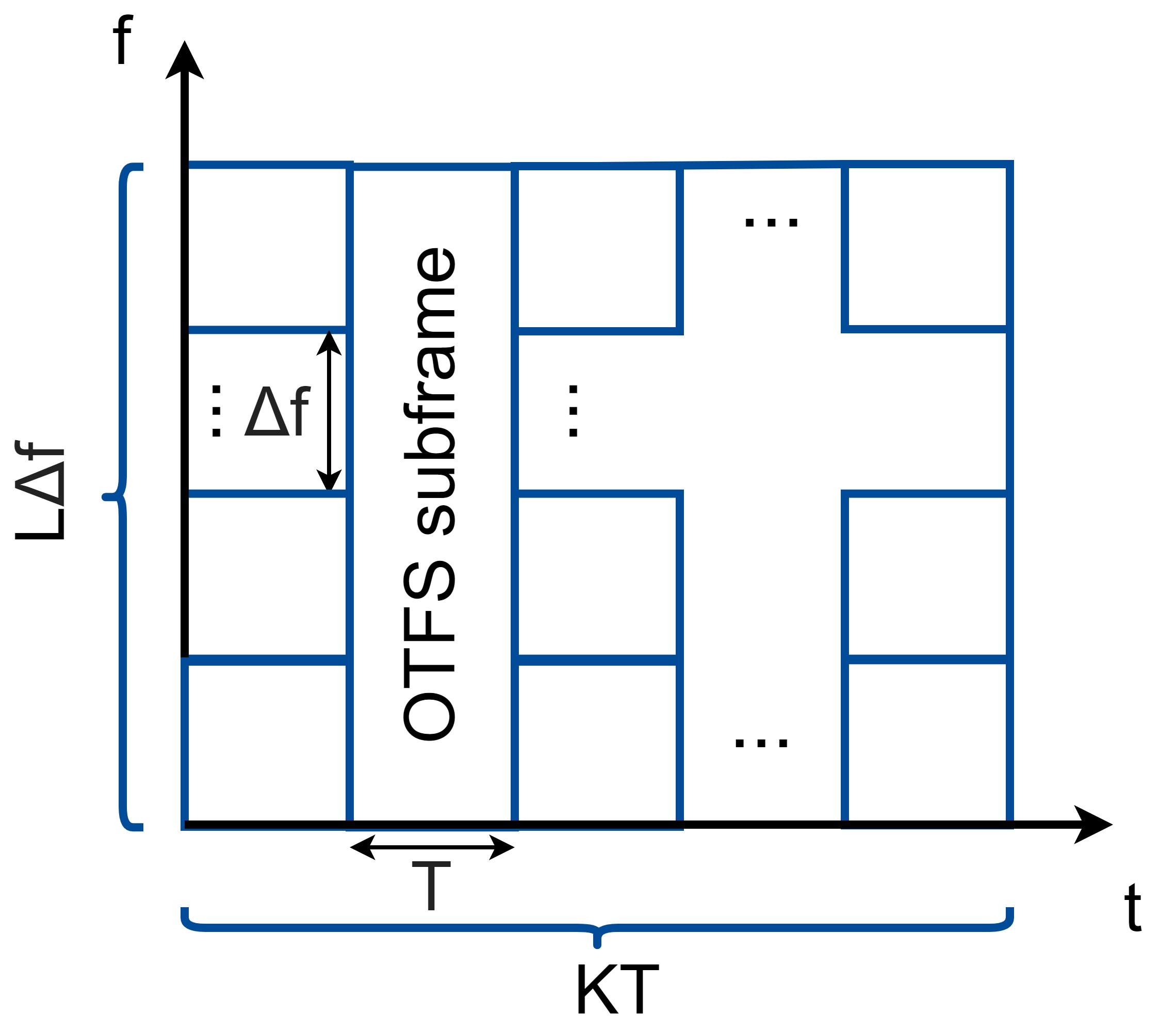}
         \caption{Before adding CP}
         \label{fig:y equals x}
     \end{subfigure}
     \hfill
     
     \begin{subfigure}[b]{\textwidth}
         \centering
         \includegraphics[width=\textwidth]{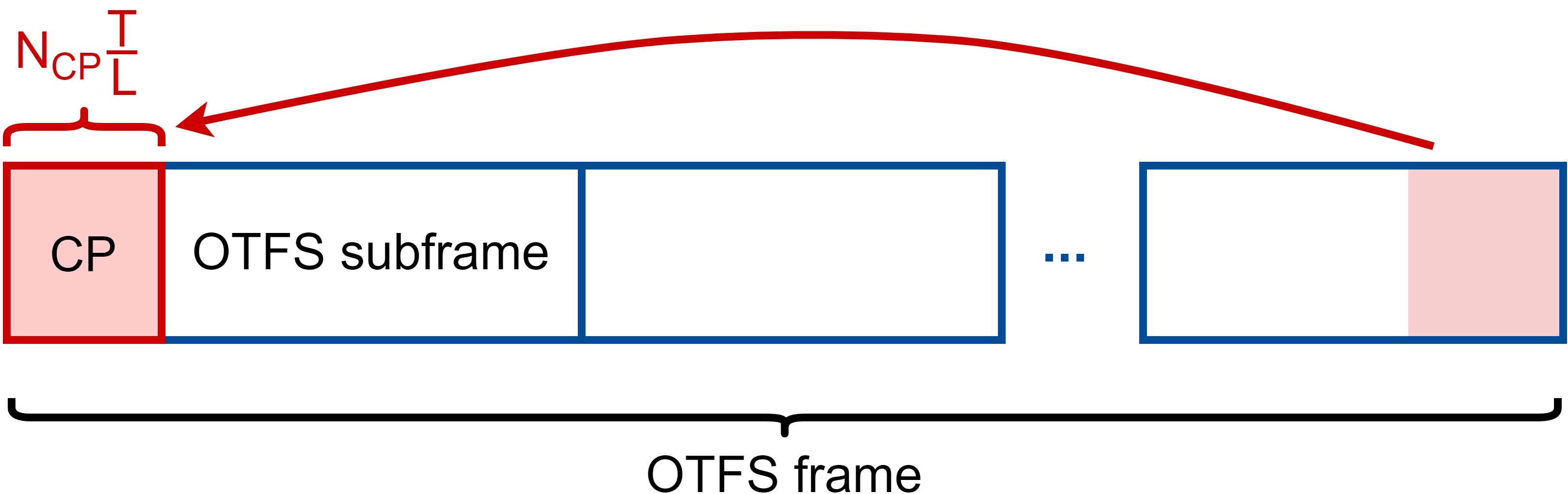}
         \caption{After adding CP}
         \label{fig:three sin x}
     \end{subfigure}
\caption{An OTFS frame structure}
\label{fig:otfs-frame}
\end{figure}

\addtolength{\topmargin}{0.05in}
\begin{figure*}
\centering
\includegraphics[width=\textwidth]{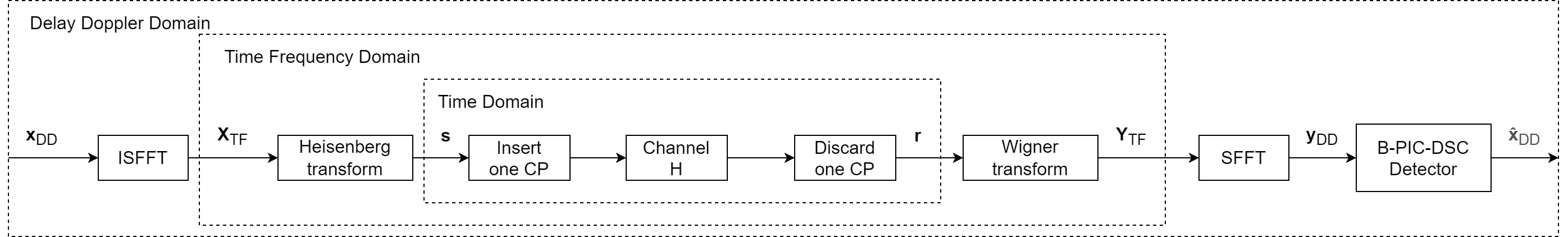}
\caption{The system model of OTFS modulation scheme}
\label{fig:sys-mod-general}
\end{figure*}

We consider an  OTFS system, illustrated in Fig. \ref{fig:sys-mod-general}. The transmitter and receiver are equipped with a single antenna. The channel state information (CSI) is assumed to be known in the receiver side. In the following, we discuss the detail of the OTFS transmitter, channel,  and receiver.  
%In the following, we use subscripts $(\cdot)_{\rm DD}$ and $(\cdot)_{\rm TF}$ to indicate that the variables are in DD and TF domains, respectively.

%% <Created-Terms>
% $M$-QAM, TF, DD, ISFFT
\subsection{OTFS Transmitter}

In the transmitter side, we first map the information binary sequences into the $M$-ary quadrature amplitude modulation ($M$-QAM) symbols. The constellation set of the $M$-QAM symbols is denoted as $\Omega$, where the constellation size is $M$. The  symbols, $\qX_{\rm DD}[l, k] \in \mathbb{C}^{L \times  K}$, are assigned in the delay Doppler (DD) domain, where $l=0,\cdots,L-1$ and $k=0,\cdots,K-1,$ are the indices of discretized delay and Doppler shifts, respectively. 
We transform the symbols from the DD domain into the time-frequency (TF) domain, as illustrated in Fig. \ref{fig:sys-mod-general}, by using the inverse symplectic finite Fourier transform (ISFFT)\cite{janssen1988zak}. Here, the TF domain is discretized to $L$ by $K$ grids with uniform intervals $\Delta f$ (Hz) and $T=1/\Delta f$ (seconds), respectively. Therefore, the sampling time is $\frac{T}{L}$.
The TF domain samples $\qX_{\rm TF} \in \mathbb{C}^{L\times K}$ in an OTFS frame, which occupies the bandwidth of $L\Delta f$ and the duration of $KT$, is given as follows:
\begin{equation}
\qX_{\rm TF} = \qF_L\qX_{\rm DD}\qF_K^{\qH},
\label{eq:sysmod-tx-isfft}
\end{equation}
where $\qF_L \in \mathbb{C}^{L\times L}$ and $\qF_K^{\qH} \in \mathbb{C}^{K\times K}$ are the discrete Fourier transform (DFT) and inverse DFT (IDFT) matrices\footnote{The $(p,q)$th entries of $N$-points DFT and its inverse are $(\frac{1}{\sqrt{N}}e^{-j2\pi pq/N})_{p,q=0,\cdots,N-1}$ and $(\frac{1}{\sqrt{N}}e^{j2\pi pq/N})_{p,q=0,\cdots,N-1}$.}, respectively. The (discrete) Heisenberg transform\cite{hadani2017orthogonal} is then applied to convert the TF domain samples into the  time domain, referred to as the transmitted signal, i.e.,
\begin{align}
\qs &= {\sf vec}(\qG_{\rm tx}\qF^{\qH}_L\qX_{\rm TF}) \notag \\ 
&={\sf vec}(\qG_{\rm tx}\qF^{\qH}_L    \qF_L\qX_{\rm DD}\qF_K^{\qH})  \notag \\ 
& = {\sf vec}(\qG_{\rm tx}\qX_{\rm DD}\qF_K^{\qH}) \notag\\
& = (\qF^{\qH}_K\otimes \qG_{\rm tx})\qx_{\rm DD},
\label{eq:sysddmod-tx-heisenberg}
\end{align}
where $\qs \in \mathbb{C}^{KL\times 1}$ is the vector of the transmitted signal, $\qG_{\rm tx} = {\sf diag}\left[g_{\rm tx}(0),g_{\rm tx}(T/L),\cdots,g_{\rm tx}((L-1)T/L)  \right] \in \mathbb{C}^{L\times L}$, 
 $\sf diag[\cdot]$ denotes the operation to diagonalize a vector, $g_{\rm tx}(t)$ is a rectangular waveform \cite{mecklenbrauker1989tutorial}, and $\qx_{\rm DD} \in \mathbb{C}^{KL \times 1}$ is the vectorization of $\qX_{\rm DD}$, by using the Kronecker product rule\footnote{A matrix multiplication is often expressed by using vectorization with the Kronecker product. That is, ${\sf vec}(ABC) = (C^T \otimes A){\sf vec}(B)$}. Note that we follow the OTFS model, considered in \cite{raviteja2018practical}, where the cyclic prefix (CP) is only inserted in the beginning of the OTFS frame,  which not only mitigates the CP overhead but also tremendously increases the spectral efficiency.
Hence, for each OTFS frame, the time duration after adding CP is $KT + N_{\rm cp}\frac{T}{L}$ and the number of samples is $KL + N_{\rm cp}$, where $ N_{\rm cp}$ is equal to the index of the maximum delay. The OTFS frame structure is shown in Fig. \ref{fig:otfs-frame}.
%, where each OTFS frame contains of $K$ number of OTFS subframes and the CP.

%We can then rewrite \eqref{eq:sysddmod-tx-heisenberg} in vector form as
%\begin{equation}
%\qs = {\sf vec}(\qS) = (\qF^{\qH}_K\otimes \qG_{\rm tx})\qx_{\rm DD},
%\label{eq:sysddmod-tx-vec}
%\end{equation}

\subsection{OTFS Wireless Channel}
The OTFS wireless channel is a time-varying multipath channel, represented by the impulse responses in the delay Doppler domain, i.e.,
\begin{equation}
h(\tau, v) = \sum_{i=1}^P h_i \delta(\tau - \tau_i)\delta(v - v_i)
\end{equation}
where $\delta(\cdot)$ is the Dirac delta function, $h_i$ denotes the $i$-th path gain, and $P$ is the total number of paths. These paths come with different delay and/or Doppler characteristics, where each of them represents the channel between moving reflectors/transmitter and the receiver. Therefore, 
%that have different delay and/or Doppler characteristics, so that the reflectors that have the same delay and Doppler characteristics are assumed to be in the same path.
the delay and Doppler shifts are given as 
\begin{equation}
\tau_i = l_i\frac{T}{L},\quad v_i = (k_i  )\frac{\Delta f}{K},
\end{equation}
respectively. Here, the integers $l_i \in [0,  l_{max}]$ and $k_i \in [-k_{max}, k_{max}]$ denote the indices of the delay and Doppler shifts,  where $l_{max}$ and $k_{max}$ are the indices of the maximum delay and maximum Doppler shifts among all channel paths. 
%Moreover, $-\frac{1}{2} < \kappa_i \leq \frac{1}{2}$ denotes the fractional Doppler.
% From  \cite{fish2013delay, raviteja2018low, raviteja2018practical}, we learn that the fractional Doppler can be handled by adding virtual integer taps in the DD channel. Therefore, to simplify the system model, we set $\kappa_i = 0$ in this work.  
% We define
%$\bold{\theta} \triangleq [\tau_1, \cdots, \tau_P]^T$ and $\bold{\psi} \triangleq [k_1,\cdots,k_P]^T$. The ranges of the delay and Doppler shift indice are [$0$, ${\sf max}(\theta)$] and [$-{\sf max}(\psi), {\sf max}(\psi)$], respectively. The largest elements in vectors $\bold{\theta}$ and $\bold{\psi}$, corresponding to the largest delay and Doppler values from all paths, are given as ${\sf max}(\theta)$ and ${\sf max}(\psi)$.
%Note that $-\frac{1}{2} < \kappa_i \leq \frac{1}{2}$ denotes the fractional Doppler. 

%
%While fractional Doppler shift $k_i???$ exist, it is unnecessary since the sampling interval $\frac{T}{L}$ in common wide-band systems provides a moderate resolution of delay taps\cite{tse2005fundamentals}. To the say the least of it, even if fractional delay and Doppler frequency shifts are obvious and considerable, those shifts can be solved by virtual integer taps insertion technique in the DD domain\cite{fish2013delay, raviteja2018low, raviteja2018practical}. Therefore, our results can be scaled into the general scenarios where fractional Doppler offsets exist.

\subsection{OTFS Receiver}

The received signal is obtained from sending the transmitted signal $\qs$ over the channel  $h(\tau,v)$, which includes the delay and Doppler terms. We discard the CP by removing the first $N_{\rm cp}$ samples from a received OTFS frame, and thus the time domain received signal is given as\cite{guillaud2003channel} 
\begin{equation}
r(n) =  \sum^P_{i} h_i e^{j2\pi \frac{k_i(n-l_i)}{KL}} s([n - l_i]_{KL}) + w(n),
\label{eq:sysmod-r(n)-scalar}
\end{equation}
where $n=0,\cdots, KL - 1$. The sampling frequency is $L\Delta f$ \cite{raviteja2018practical}. 
We can then rewrite \eqref{eq:sysmod-r(n)-scalar} in a matrix-vector form as
\begin{equation}
\qr = \qH\qs + \qw,
\label{eq:sysmod-r(n)}
\end{equation}
where $\qw$ is the independent and identically distributed (i.i.d.) white Gaussian noise that follows $\mathcal{N}(\bold{0}, \sigma^2 \qI)$, $\sigma^2$ is the variance of the noise, and $\qH = \sum_{i=1}^P h_i \qI_{KL}(l_i) \qDelta({k_i})$, $\qI_{KL}(l_i)$ denotes a $KL\times KL$ matrix obtained by circularly left shifting the columns of the identity matrix by $l_i$, for example when $l_i =1$,
\[
\qI_{KL}(1) = \begin{bmatrix}
0 & \cdots & 0 & 1\\
1 & \ddots & 0 & 0\\
\vdots & \ddots & \ddots & \vdots\\
0 & \cdots & 1 & 0\\
\end{bmatrix}	.
\]
Furthermore,  $\qDelta$ is the $KL \times KL$ diagonal matrix from the Doppler shifts, i.e., 
%\begin{equation}
\[
\qDelta({k_i}) = {\sf diag}\left[e^{\frac{j2\pi k_i(0)}{KL}}, e^{\frac{j2\pi k_i(1)}{KL}}, \cdots, e^{\frac{j2\pi k_i(KL - 1)}{KL}}\right].
\]
%\end{equation}
%$
%\qDelta({k_i}) = \begin{bmatrix}
%e^{\frac{j2\pi k_i(0)}{KL}} & 0 & \cdots & 0\\
%0 & e^{\frac{j2\pi k_i(1)}{KL}} & \ddots & \vdots\\
%\vdots & \ddots & \ddots &  0 \\ 
%0 & \cdots & 0 & e^{\frac{j2\pi k_i(KL - 1)}{KL}}
%\end{bmatrix}	.
%$\\
Note that the matrices $\qI_{KL}(l_i)$ and $\qDelta({k_i})$ describe the delay and Doppler shifts in \eqref{eq:sysmod-r(n)-scalar}, respectively. We define a matrix  \[
\qR \triangleq  \begin{bmatrix}
r(0) & r(L) & \cdots & r((K-1)L)\\
r(1)  & r(L+1) & \cdots & r((K-1)L + 1)\\
\vdots & \vdots & \cdots &  \vdots \\ 
r(L-1) & r(2L-1) & \cdots & r(KL-1)
\end{bmatrix}	.
\]

As shown in Fig. \ref{fig:sys-mod-general}, we convert the received signal into the TF domain by applying the Wigner transform \cite{raviteja2018practical},  which yields 
\begin{equation}
\qY_{\rm TF} = \qF_L\qG_{\rm rx}\qR,
\label{eq:sysddmod-rx-wigner}
\end{equation}
where $\qG_{\rm rx} = {\sf diag}\left[g_{\rm rx}(0),g_{\rm rx}(T/L),\cdots,g_{\rm rx}((L-1)T/L)  \right] \in \mathbb{C}^{L\times L}$ and $g_{\rm rx}(t)$ is the rectangular waveform in the receiver. 
 We then use the symplectic finite Fourier transform (SFFT)\cite{janssen1988zak} to obtain the received signal in the DD domain, i.e.,
\begin{align}
\qY_{\rm DD}  &=\qF^{\qH}_L \qY_{\rm TF} \qF_K \notag \\
&= \qF^{\qH}_L \qF_L\qG_{\rm rx}\qR \qF_K \notag \\
&= \qG_{\rm rx}\qR\qF_K.
\label{eq:sysmod-rx-sfft}
\end{align}
By following the vectorization with Kronecker product rule, we can rewrite \eqref{eq:sysmod-rx-sfft} as
\begin{align}
\qy_{\rm DD} &= {\sf vec}(\qY_{\rm DD}) \notag \\
&= {\sf vec}(\qG_{\rm rx}\qR\qF_K) \notag \\
%&= (\qF_K^{\rm T} \otimes \qG_{\rm rx}){\sf vec}(R) \notag\\
%&= (\qF_K^{\rm T}  \otimes \qG_{\rm rx})\qr \notag \\
& = (\qF_K \otimes \qG_{\rm rx})\qr.
\label{eq:sysmod-rx-vec}
\end{align}
By first substituting $\qs$ in \eqref{eq:sysddmod-tx-heisenberg} into \eqref{eq:sysmod-r(n)} and using the result to replace $\qr$ in  \eqref{eq:sysmod-rx-vec}, we obtain
\begin{align}
\qy_{\rm DD} = \qH_{\rm eff}\qx_{\rm DD} + \tilde{\qw},
%&= (\qF_K \otimes \qG_{\rm rx})\qH(\qF^{\qH}_K \otimes \qG_{\rm tx})\qx_{\rm DD} + (\qF_K \otimes \qG_{\rm rx})\qw \notag \\
%&= \qH_{\rm eff}\qx + \tilde{\qw},
\label{sysddmod-y=hx+w}
\end{align}
% $\qH_{\rm eff} = (\qF_N \otimes \qI_L)\qH(\qF^{\qH}_N \otimes \qI_L)$
% $\tilde{\qw} = (\qF_N \otimes \qI_L)\qw$ 
where $\qH_{\rm eff} = (\qF_K \otimes \qG_{\rm rx})\qH(\qF^{\qH}_K \otimes \qG_{\rm tx})$ and $\tilde{\qw} =(\qF_K \otimes \qG_{\rm rx})\qw$ denote the effective channel and noise in the DD domain, respectively.  Here, $\tilde{\qw}$ is an i.i.d. Gaussian noise, since $\qF_K \otimes \qG_{\rm rx}$ is a unitary orthogonal matrix \cite{hadani2017orthogonal, raviteja2018practical}.
%, since $(\qF_K \otimes \qG_{\rm rx})$ is a unitary transformation that does not change the noise distribution.
%$\qG_{\rm tx}$ and $\qG_{\rm rx}$ are rectangular waveforms  \cite{hadani2017orthogonal, raviteja2018practical} which are equal to the identity matrix with size $L$, since they are used to select $L$ samples each time.
%(i.e., $\qG_{\rm tx} = \qG_{\rm rx} = \qI_L$) 
%Therefore we can re-express $\qH_{\rm eff} = (\qF_K \otimes \qI_L)\qH(\qF^{\qH}_K \otimes \qI_L)$ and $\tilde{\qw} = (\qF_K \otimes \qI_L)\qw$.
%while the bi-orthogonal waveform does not exist in reality and out-of-band power reduction causes nonuniform channel gains for transmitted QAM symbols \cite{hadani2017orthogonal, raviteja2018practical} \red{[cite Viterbo's reduced CP paper]}. 
%Moreover, $\qF_K \otimes \qI_L$ is a unitary transformation that does not change the distribution of ${\qw}$.
% variance of the distribution \cite{freedman1980empirical, yuan2020iterative}. Hence, $\tilde{\qw}$ can be rendered as a Gaussian distribution with mean 0 and variance $\sigma_w^2$. $\qG_{\rm tx}$ and $\qG_{\rm rx}$ are rectangular waveforms (i.e., $\qG_{\rm tx} = \qG_{\rm rx} = \qI_L$) that guarantees a uniform channel gain to all symbols, while the bi-orthogonal waveform does not exist in reality and out-of-band power reduction causes nonuniform channel gains for transmitted QAM symbols \cite{hadani2017orthogonal, raviteja2018practical} \red{[cite Viterbo's reduced CP paper]}. 

\section{B-PIC-DSC OTFS Detector}

\begin{figure*}
\centering
\includegraphics[width=0.68\textwidth]{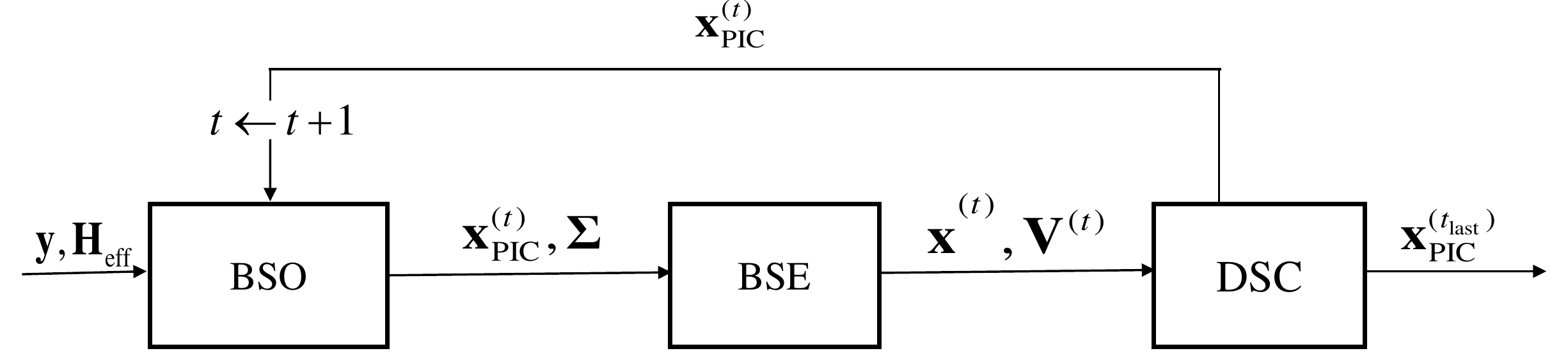}
\caption{B-PIC-DSC Detector}
\label{fig:B-PIC-DSC Detector}
\end{figure*}

In this section, we describe the development of the  B-PIC-DSC detector\cite{kosasih2020linear} for an OTFS system. The structure of the B-PIC-DSC detector is illustrated in  Fig. \ref{fig:B-PIC-DSC Detector}. It consists of three modules: Bayesian symbol observation (BSO), Bayesian symbol estimation (BSE), and DSC modules. Note that in the following discussion we consider the DD domain received signal, where the expression is given in \eqref{sysddmod-y=hx+w}. In the following, we omit the subscripts $\rm {DD}$ and $\rm eff$ for notational simplicity.
%that computes the pdfs of the transmitted symbols from the  received signals by using  the matched filter based PIC scheme; a BSE module that obtains the Bayesian symbol estimates based on the computed pdfs; and a DSC module that refines the transmitted symbol estimates by using the outputs of the BSE module and returns the refined symbols to the BSO module. 

\subsection{Bayesian Symbol Observation}

In the BSO module,  the transmitted symbols  $\qx$ in \eqref{sysddmod-y=hx+w} is treated as a random vector.  The posterior function of $\qx = [x_1, \cdots, x_q, \cdots, x_{KL}]$, given the received signals $\qy = [y_1, \cdots, y_c, \cdots, y_{KL}]$, is expressed as 
\begin{equation}\label{Bayesian_rule}
p(\qx|\qy) = \frac{p(\qy|\qx) p(\qx)}{p(\qy)},
\end{equation}
where $p(\qy|\qx) =\mathcal{N}\left( \qy, \qH\qx; \sigma^2\qI \right)$. 
%ince the transmitted symbols are uniformly distributed, $p(\qx|\qy)$  in \eqref{Bayesian_rule} can be simplified as 
%\begin{equation}\label{dist_of_x_y}
%p(\qx|\qy) \propto \mathcal{N}\left( \qy, \qH\qx; \sigma^2\qI \right).  
%\end{equation}
 Following\cite{kosasih2020linear}, we iteratively approximate the posterior probability $p(\qx|\qy)$ by a product of independent Gaussian functions  expressed as
\begin{equation}\label{iid_assumption}
p(\qx|\qy)   \approx \prod_{q=1}^{KL} \underbrace{\mathcal{N}\left( x_q, x_{{\rm PIC},q}^{(t)}; \Sigma_q^{(t)} \right)}_{\hat{p}^{(t)}(x_q|\qy)}.
\end{equation}
Here, $x_{{\rm PIC},q}^{(t)}$  is the soft estimate  of $x_q$ in iteration $t$, obtained from  the matched filter based PIC scheme, given as follows:
\begin{equation}\label{eA1_a02}
x_{{\rm PIC},q}^{(t)} = \frac{\qh_q^{H} \left( \qy-\qH \qx_{{\rm PIC} \backslash q}^{(t-1)} \right)}{\| \qh_q \|^2},
\end{equation} 
where $\qh_q$ is the $q$-th column of matrix $\qH$ and \[  \qx_{{\rm PIC} \backslash q}^{(t-1)}   = \left[x_{{\rm PIC}, 1}^{(t-1)}  , \dots, x_{{\rm PIC}, q-1}^{(t-1)} , 0, x_{{\rm PIC}, q+1}^{(t-1)}, \dots, x_{{\rm PIC}, KL}^{(t-1)} \right]^{\rm T} \] is the estimated symbols in the $(t-1)$-th iteration. Note that we use the MMSE scheme to produce the initial value of the $q$-th symbol estimate, that is, 
\begin{equation}\label{MMSE_init}
 x_{{\rm PIC}, q}^{(0)} = \left(\qh_q^{H} \qh_q + \sigma^2 \right)^{-1}\qh_q^{H} \qy.
\end{equation}
\begin {Remark}
 The MMSE initialization given in \eqref{MMSE_init} is necessary to improve the performance of the maximum ratio combining (MRC) based B-PIC-DSC detector proposed in\cite{kosasih2020linear}, especially in the presence of high interference. Such high interference is encountered in the OTFS systems when there is a large number of moving reflectors, namely, high ICI.
% when there is a large  number of interference. Such high interference is encountered in rich scattering OTFS systems where the ICI and ISI are high.
\end {Remark}
The variance $ \Sigma_q $ of the  $q$-th PIC symbol estimate  is approximated in\cite{kosasih2020linear} as
\begin{align}\label{eA1_a01}
\Sigma_q &   = \frac{\sigma^2}{\sum_{c=1}^{KL}h_{c,q}^* h_{c,q}}.
\end{align}
  The approximated posterior functions,  $\hat{p}^{(t)}(x_q|\qy)={\cal{N}}\left( x_q,x^{(t)}_{{\rm PIC},q}; {\Sigma}^{(t)}_q \right), q=1, \dots, KL,$  are then forwarded to the BSE module, as shown in Fig. \ref{fig:B-PIC-DSC Detector}.
  
\subsection{Bayesian Symbol Estimator}
      
In the BSE module, we compute the Bayesian symbol estimate of the $q$-th symbol by using $\hat{p}^{(t)}(x_q|\qy)$ obtained from the BSO module.  
Based on the factorization of $p(\qx|\qy)$  in \eqref{iid_assumption}, we infer the symbol estimate $\hat{x}_q^{(t)}$ by using the maximum a posteriori criterion, given as 
\begin{flalign}\label{MAP_component}
\hat{x}_q^{(t)} &= \arg \max_{a \in \Omega} \hat{p}^{(t)}(x_q=a|\qy).
\end{flalign}
Note that the MAP criterion in \eqref{MAP_component}  has a linear complexity since the inference is performed for each $q$-th symbol estimate. 
The Bayesian symbol estimate and its variance  are respectively given as
\begin{equation}\label{eA1_b01}
\hat{x}_q^{(t)} =\Ex\left[x_q \Big| x_{{\rm PIC}, q}^{(t)} ,\Sigma_q \right] =\sum_{a \in \Omega} a  \hat{p}^{(t)}{\left(x_q=a|\qy\right)}
\end{equation}
\begin{equation}\label{eA1_b02}
V_q^{(t)}=\Ex  \left[ \left| x_q  - \Ex\left[x_q \Big| x_{{\rm PIC}, q}^{(t)} ,\Sigma_q \right] \right|^{2} \right], 
\end{equation}
where $\hat{p}^{(t)}\left(x_q|\qy\right)$ is normalized so that  $\sum_{a\in \Omega}\hat{p}^{(t)}\left(x_q=a|\qy\right) =1$.
The outputs of the BSE module, $\hat{x}_q^{(t)} $ and $V_q^{(t)} $, $q= 1, \dots, K$, are then sent to the DSC module.

\subsection{Decision Statistics Combining}

The correlation between $\hat{x}_q^{(t)}$ and $\hat{x}_q^{(t-1)}$ in the B-PIC-DSC detector is low in the early iteration stages\cite{kosasih2020linear}. 
Such a feature can be exploited to increase the diversity of symbol estimates by forming decision statistics,  referred to as the DSC concept\cite{vucetic2003space}. The decision statistics consist of a linear combination of the symbol estimates in two consecutive iterations 
\begin{equation}\label{DSC}
x_{{\rm DSC},q}^{(t)} = \left( 1-\rho_{{\rm DSC},q}^{(t)} \right)  \hat{x}_q^{(t-1)}  +   \rho_{{\rm DSC},q}^{(t)}   \hat{x}_q^{(t)}
\end{equation}
%\begin{equation}\label{DSC_Var}
%V_{{\rm DSC},q}^{(t)} = \left( 1-\rho_{{\rm DSC},q}^{(t)} \right)  V_q^{(t-1)}  +   \rho_{{\rm %DSC},q}^{(t)}   V_q^{(t)}.
%\end{equation}
The weighting coefficient,
\begin{equation}\label{DSC_coef}
\rho_{{\rm DSC},q}^{(t)} =  \frac{e_q^{(t-1)}}{e_q^{(t)}+e_q^{(t-1)}}, 
\end{equation}
is determined by maximizing the signal-to-interference-plus-noise-ratio (SINR). 
Here, $e_q^{(t)}$ is defined as the instantaneous square error of the $q$-th symbol estimate, computed by using the MRC filter,
\begin{flalign}\label{DSC_error}
 e_q^{(t)}  =  \left\|\frac{\qh_q^{\qH}}{\| \qh_q\|^2} \left(  \qy - \qH \hat{\qx}^{(t)} \right)\right\|^2.
\end{flalign}
The iterative process is terminated if the following condition is satisfied,
\begin{equation}\label{eq_convergence}
 \|x_{{\rm DSC},q}^{(t)} - x_{{\rm DSC},q}^{(t-1)} \| \leq \zeta  \text{  or  } t = t_{\rm max}, 
\end{equation}
where  $\zeta$ is the minimum acceptable difference of $x^{(t)}_{{\rm DSC},q}$ in two consecutive iterations, and $t_{\rm max}$ is the maximum number of iterations. We then use $x_{{\rm DSC},q}^{(t)} $ as the input of the BSO module by assigning the value of $x_{{\rm DSC},q}^{(t)}$ to $x_{{\rm PIC},q}^{(t)}$,
\begin{equation}\label{assign}
x_{{\rm PIC},q}^{(t)} \leftarrow x_{{\rm DSC},q}^{(t)}.
\end{equation}
 %\gray{,\begin{equation}\label{Assign}x_{{\rm PIC},q}^{(t)} \leftarrow  {x}_{{\rm DSC},q}^{(t)},  \text{ and }  V_q^{(t)} \leftarrow V_{{\rm DSC},q}^{(t)},  q =1,\dots,KL.\end{equation}}
%The iteration is repeated until \eqref{eq_convergence} is zsatisfied.
%Then, the hard decision is made according to}
%\begin{equation}\label{Hard_Dec}
%x_{q}^{\rm hard} =\arg \min_{x_q \in \Omega} \| x_q - x_{{\rm PIC},q}^{(T)}  \|^2, q =1,\dots,KL
%\end{equation}

\begin{algorithm}
\caption{The B-PIC-DSC OTFS detector\label{A1} }
\label{A1}
\begin{algorithmic}[1]
\State {\textbf{Input: }$K, \Omega, \qy, \qH, \sigma^2,  V_q^{(0)} \leftarrow 1 , t_{\rm max} \leftarrow 10 $}
\State {\textbf{Output: }  $\hat{\qx}^{(T)}$}
	\For {$t=1,\dots, t_{\rm max}$}
		\For {$q=1,\dots, KL$ (or parallel execution)}
	    		\Statex \textbf{\quad\ \quad\, The BSO Module:}
	    		\If {$t=0$}
			\State {Compute \eqref{MMSE_init}}
			\EndIf
			\State Compute $x_{{\rm PIC},q}^{(t)}$  in \eqref{eA1_a02} 
			\State Compute the variance ${\Sigma}_q$  in \eqref{eA1_a01}
			\Statex \textbf{\quad\ \quad\, The BSE Module:}
			\State Compute  the Bayesian symbol estimate $\hat{x}_q^{(t)}$  in \eqref{eA1_b01}
			\State  Compute the Bayesian variance   $V_q^{(t)} $ in \eqref{eA1_b02}
			\Statex \textbf{\quad\ \quad\, The DSC Module:}
			\State Compute $e_{q}^{(t)} $  in \eqref{DSC_error} 
			\State Compute  $\rho_{{\rm DSC},q}^{(t)} $  in \eqref{DSC_coef} 
			\State Compute  $x_{{\rm DSC},q}^{(t)} $  in \eqref{DSC} 
			%\State Compute $V_{{\rm DSC},q}^{(t)}$ in \eqref{DSC_Var}  
			\State Compute \eqref{assign}  
		\EndFor
		\If {$ \|x_{{\rm DSC},q}^{(t)} - x_{{\rm DSC},q}^{(t-1)} \| \leq 10^{-4} $}
			\State {break}
		\EndIf
	\EndFor
		\State $t_{\rm last}\leftarrow t$ 
%\State Calculate hard symbol estimates from $\hat{\qx}^{(T)}$
%\State \textbf{Return:}
\end{algorithmic}
\end{algorithm}
The complete pseudo-code of the B-PIC-DSC OTFS detector is shown in Algorithm \ref{A1}.

\begin{table}
  \begin{tabular}{| c| c|}
  \hline
 Detector 			&     	Complexity		\\ 
     	 \hline
  \hline
    		MMSE OTFS 			& 		$\mathcal{O} (K^3L^3)$ 				\\
  \hline
    		MP OTFS 			& 		$\mathcal{O} (KLPMt_{\rm last})$ 				\\
     	 \hline
    		AMP OTFS		& 		$\mathcal{O}(K^2L^2t_{\rm last})$ 	 			\\
    		\hline
     	UTAMP OTFS & 		$\mathcal{O} (L^2K + LK M)$   			\\
     	 \hline
     	B-PIC-DSC OTFS 		& 		$\mathcal{O}(K^3L^3 + K^2L^2t_{\rm last})$ 	 		\\
     	 \hline
  \end{tabular}  \label{T1}
  \caption{Computational complexity comparison}
\end{table}

\section{Complexity Analysis}\label{Complexity}

In this section, we analyze the complexity of the proposed B-PIC-DSC OTFS detector. Algorithm 1 specifies that the B-PIC-DSC OTFS detector performs matrix vector multiplications in \eqref{eA1_a02}, \eqref{eA1_a01}-\eqref{DSC_error}, at each iteration and therefore the cost is $\mathcal{O} (K^2L^2t_{\rm last})$, where $t_{\rm last}$ denotes the number of iterations. The matrix inverse operation, performed in the first iteration, given in \eqref{MMSE_init}, requires $\mathcal{O}(K^3L^3 )$ computational complexity. Hence the overall complexity of the B-PIC-DSC OTFS detector is $\mathcal{O}(K^3L^3 + K^2L^2t_{\rm last})$. This complexity is similar to that of the MMSE OTFS detector. The computational complexity of the MP OTFS, AMP OTFS, and UTAMP OTFS detectors are tabulated in Table 1. Although the proposed B-PIC-DSC OTFS detector has a higher complexity compared to the other detectors, this comes with a significant BER performance gain in the presence of strong ICI, resulted from the presence of a large number of reflectors, as will be discussed in the next section. 
%Besides, \blue{$KL=56$ in the smallest LTE resource block usage considered in our experiment\cite{surabhi2019diversity}}.
%The B-PIC-DSC OTFS detector requires more computing power to achieve a high detection performance, especially in the rich scattering environment.

\section{Numerical Results}\label{Sim_configs}

\begin{figure}
\centering
\subfloat[P=6]
{\includegraphics[width=0.51\textwidth]{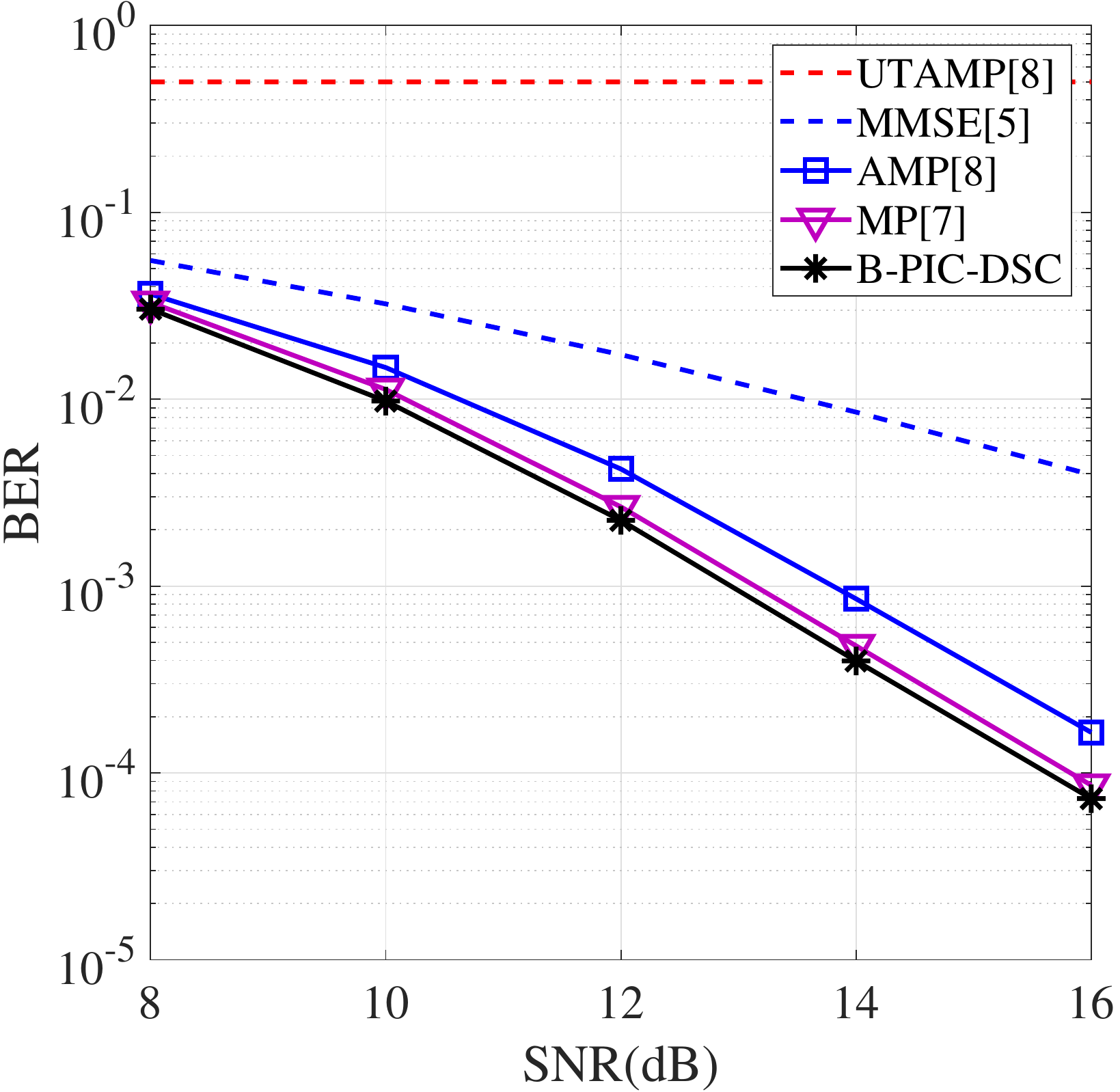}\hfill}
\subfloat[P=10]
{\includegraphics[width=0.51\textwidth]{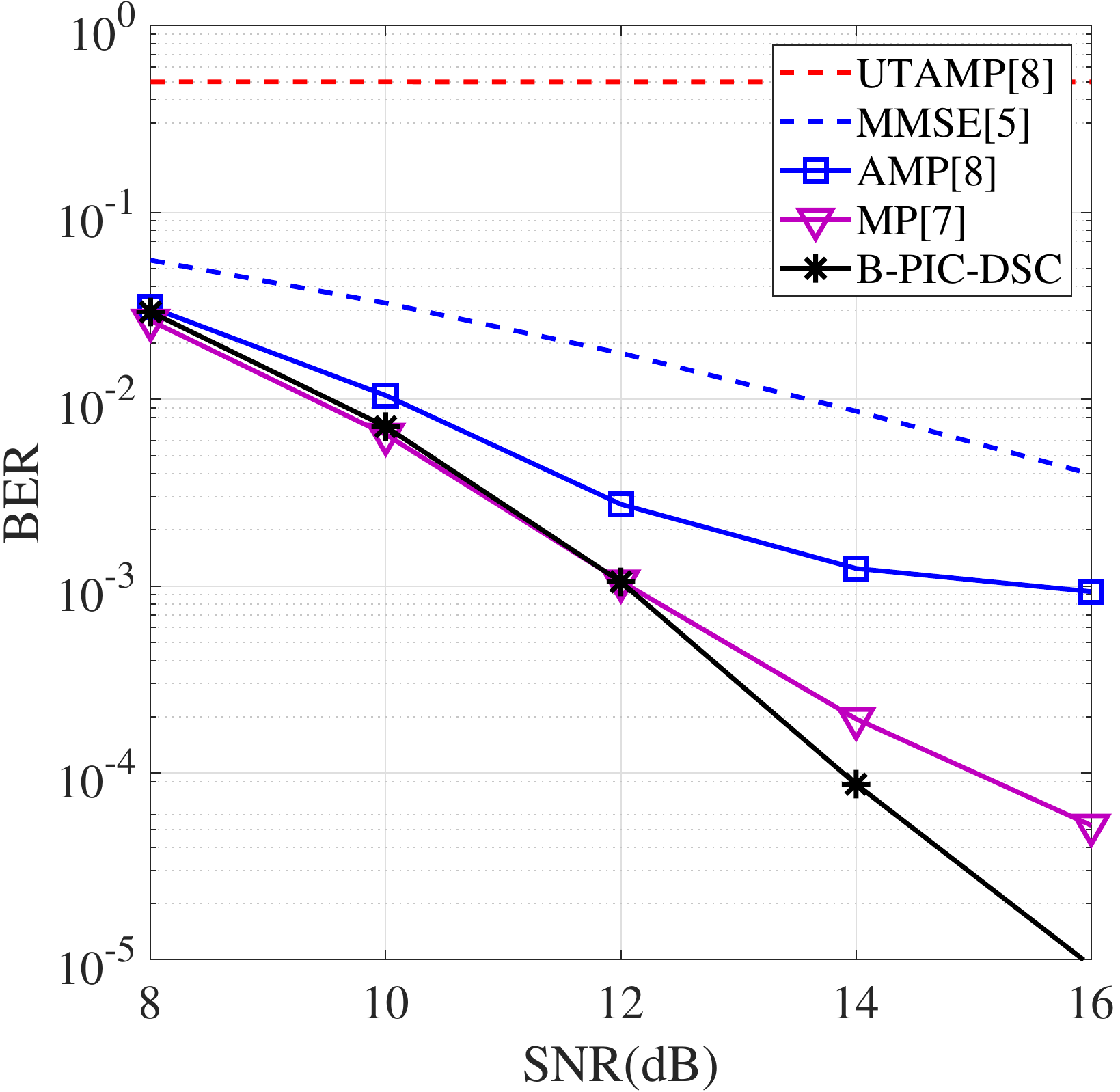}\hfill}
\caption{A moderate number of mobile reflectors with $k_{max}=1$}
\label{Fig:IB-PIC-DSC-Result-good}
\end{figure} 

\begin{figure}
\centering
\subfloat[P=6]
{\includegraphics[width=0.51\textwidth]{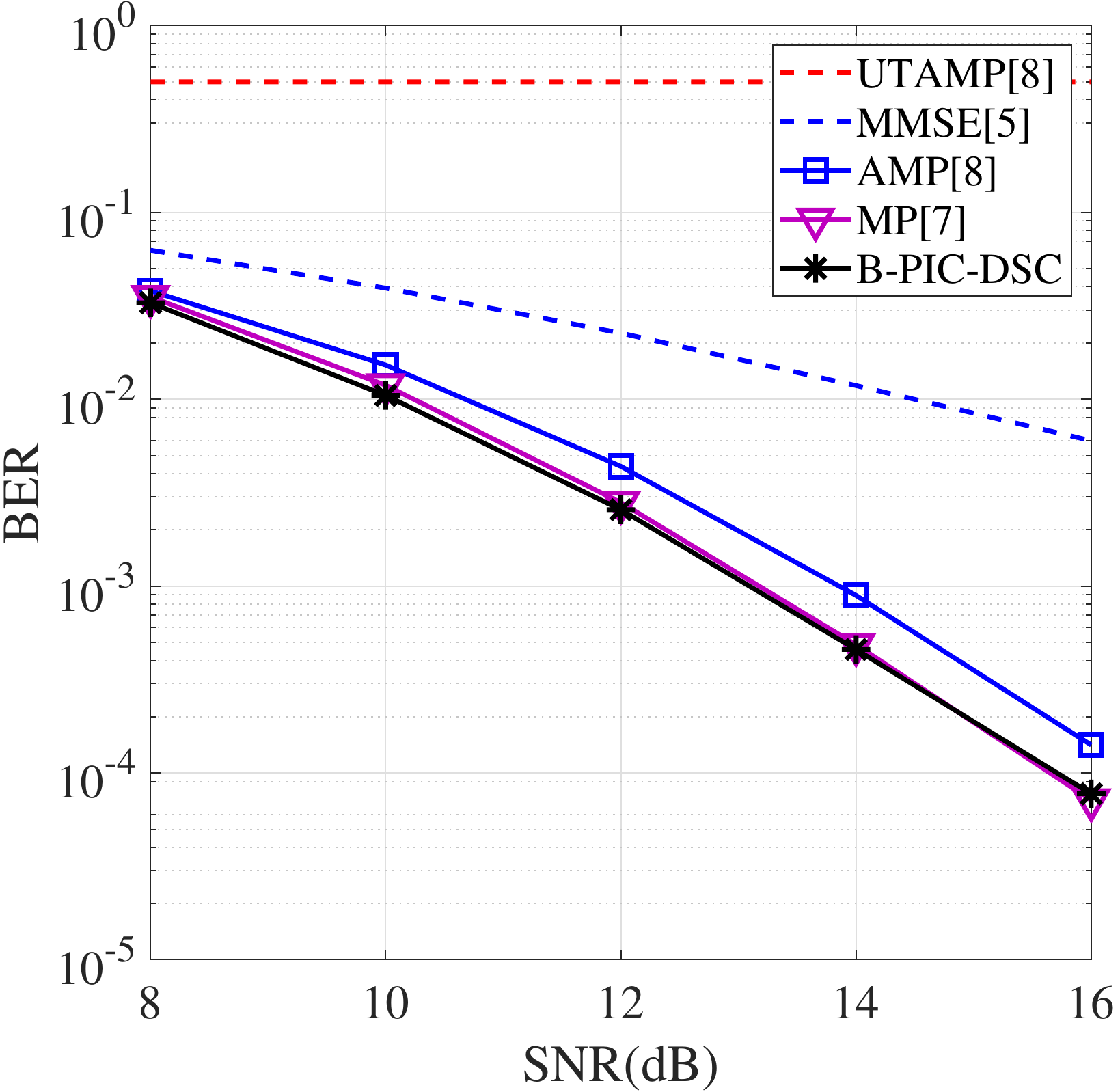}\hfill}
\subfloat[P=10]
{\includegraphics[width=0.51\textwidth]{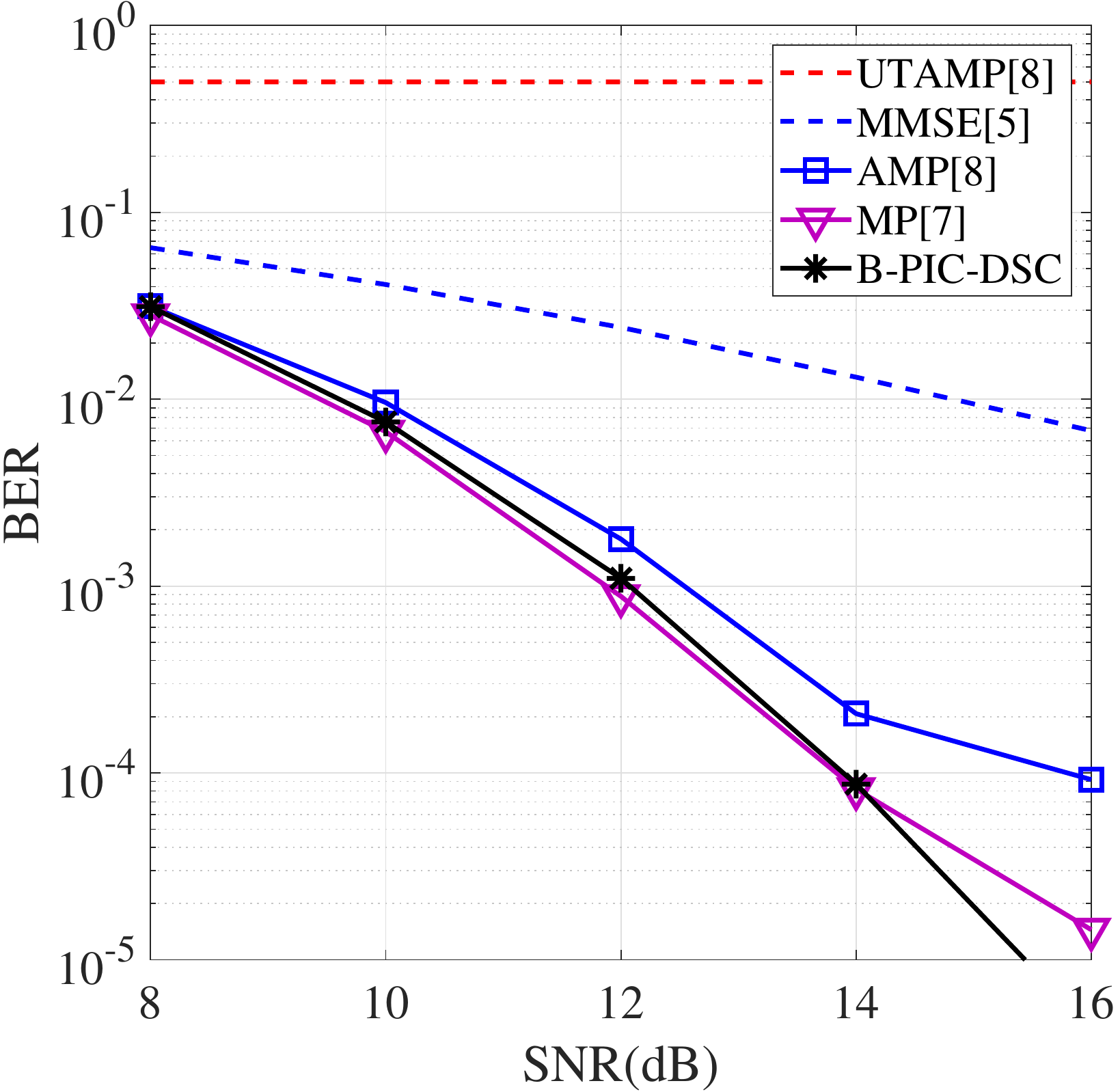}\hfill}
\caption{A moderate number of mobile reflectors with $k_{max}=3$}
\label{Fig:IB-PIC-DSC-Result-good-richscattering}
\end{figure} 

In this section, we evaluate the performance of our detector by comparing the BER of our proposed detector with that of the MMSE OTFS \cite{raviteja2018interference}, MP OTFS \cite{caire2004iterative}, AMP OTFS\cite{khumalo2016fixed}, and  UTAMP OTFS \cite{yuan2020iterative} detectors. We set $L=12$, $K=7$,  and $\Delta f=15$kHz. In the channel, the delay index is $l_i \in [1,6]$ excluding the first path ($l_i = 0$), the Doppler shift index of $i$-th path, $k_i$, is uniformly drawn from $[-1,1]$ or $[-3,3]$ and the path gain $h_i$ is independently drawn from the complex Gaussian distribution $\mathcal{N}(0, 1/P)$. The $4$-QAM modulation is employed for the simulations. The number of of the transmitted QAM symbols per OTFS frame is $KL=84$ and the number of samples in CP is $N_{\rm CP}=6$.
 
 %\red{The number of the transmitted QAM symbols, sent in an OTFS frame, equals to the number of the paths where the symbols are randomly placed in the DD grids.}

%We consider both the low and high Doppler scenario where the Doppler frequency shift index ranges are $[-1, 1]$ and $[-3, 3]$, respectively. We consider delay index $[0, 6]$ with $4$-QAM modulation. 
% to the relative velocity of $506.25$ km/h, considered in $5$G applications\cite{surabhi2019diversity} and $[-3, 3]$, respectively} . We consider delay index $[0, 6]$, $4$-QAM modulation. 

We first consider two numbers of paths $P=6$ and $P=10$. The simulation results are shown in the Figs. \ref{Fig:IB-PIC-DSC-Result-good} and  \ref{Fig:IB-PIC-DSC-Result-good-richscattering}. Here, the UTAMP OTFS detector fails to achieve an acceptable detection performance. This is because the unitary transformation is not applicable in our system model. Additionally, the MP OTFS detector outperforms the AMP and MMSE detectors. Nevertheless, the B-PIC-DSC OTFS detector can achieve the lowest BER in comparison to the other counterparts.

We then increase the numbers of paths $P=14$ and $P=18$ as illustrated in Figs. \ref{Fig:IB-PIC-DSC-Result-bad} and \ref{Fig:IB-PIC-DSC-Result-bad-richscattering}. The MP based detectors can only achieve a BER performance of $10^{-4}$ at best, indicating that these detectors cannot handle a high ICI issue, in the  OTFS system. In contrast to those existing OTFS detectors, the BER of the B-PIC-DSC OTFS detector reaches a BER less than $10^{-5}$, when SNR is over $14$ dB. Therefore, we conclude that our proposed detector  significantly outperforms the state-of-the-art OTFS detectors.

\begin{figure}
\centering
\subfloat[P=14]
{\includegraphics[width=0.51\textwidth]{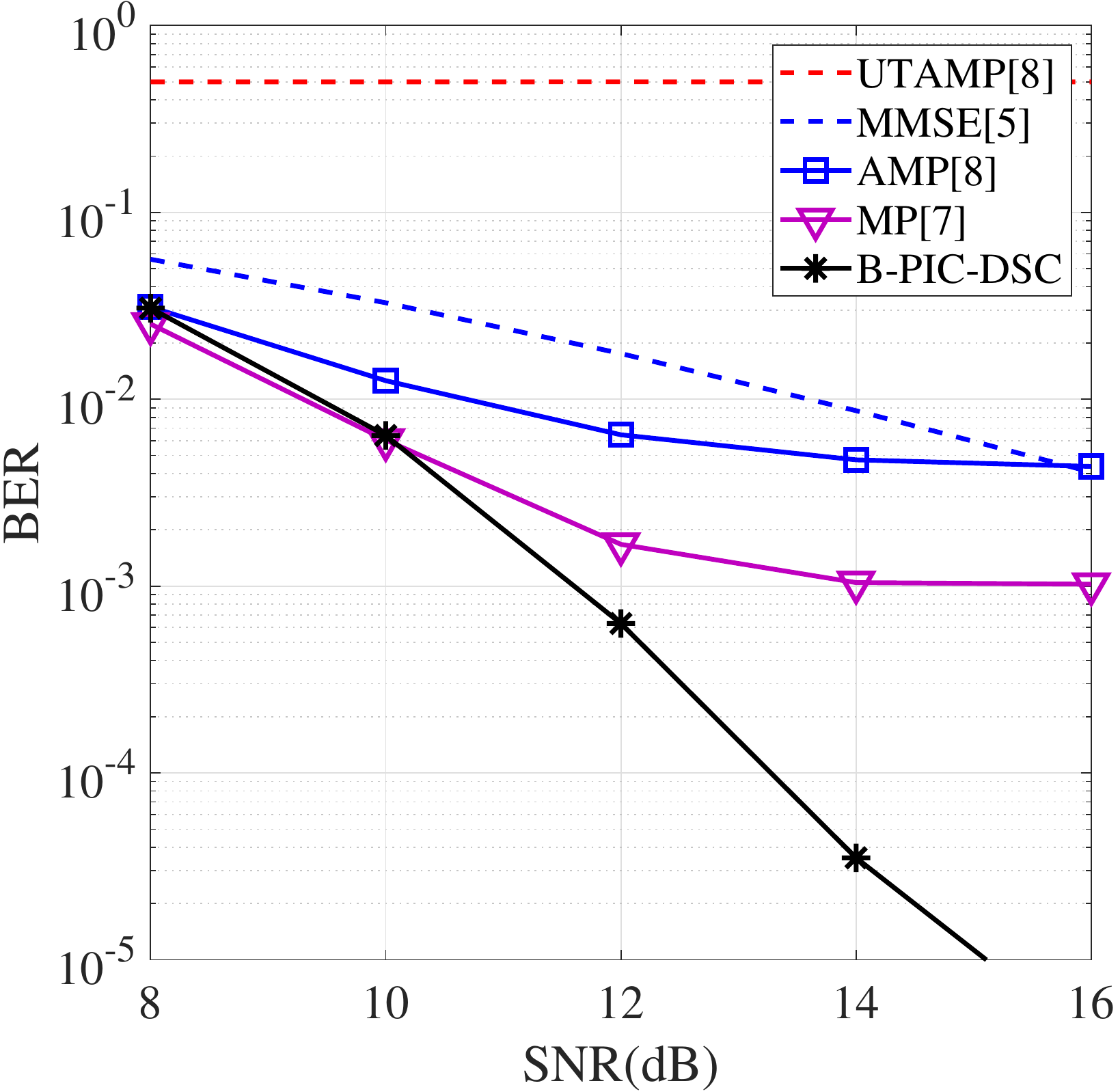}\hfill}
\subfloat[P=18]
{\includegraphics[width=0.51\textwidth]{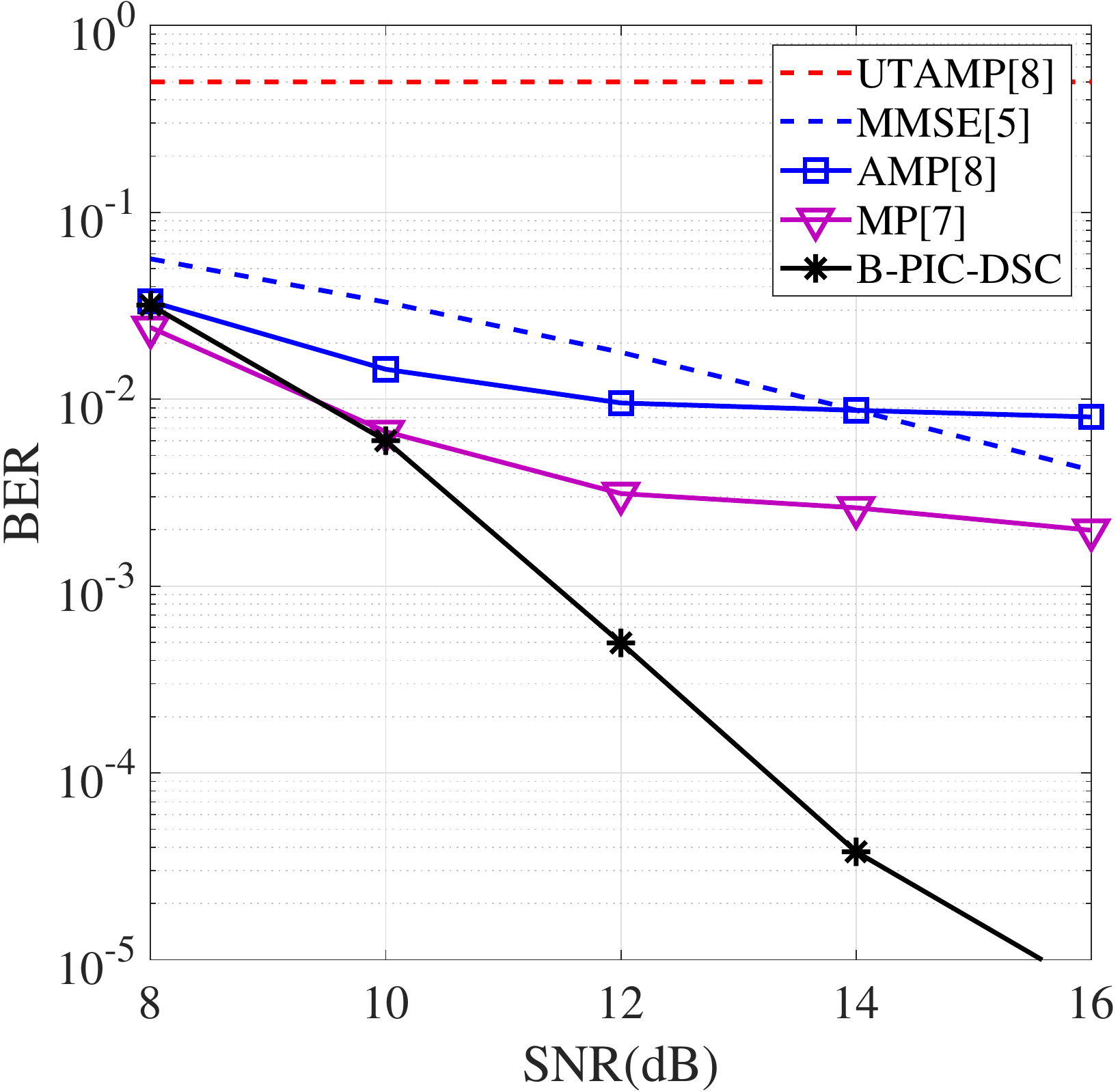}\hfill}
\caption{A large number of mobile reflectors with $k_{max} =1$}
\label{Fig:IB-PIC-DSC-Result-bad}
\end{figure}

\begin{figure}
\centering
\subfloat[P=14]
{\includegraphics[width=0.51\textwidth]{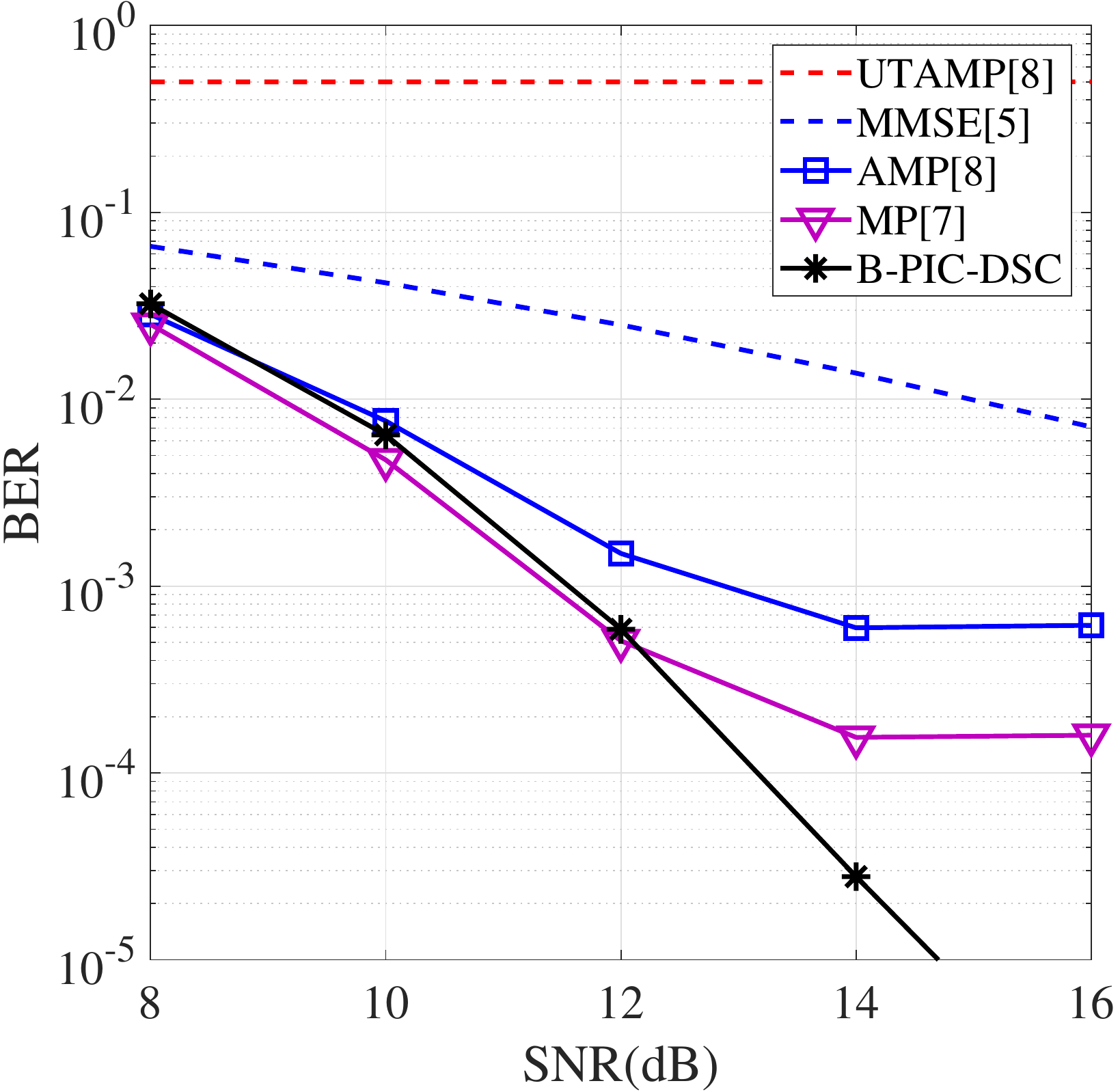}\hfill}
\subfloat[P=18]
{\includegraphics[width=0.51\textwidth]{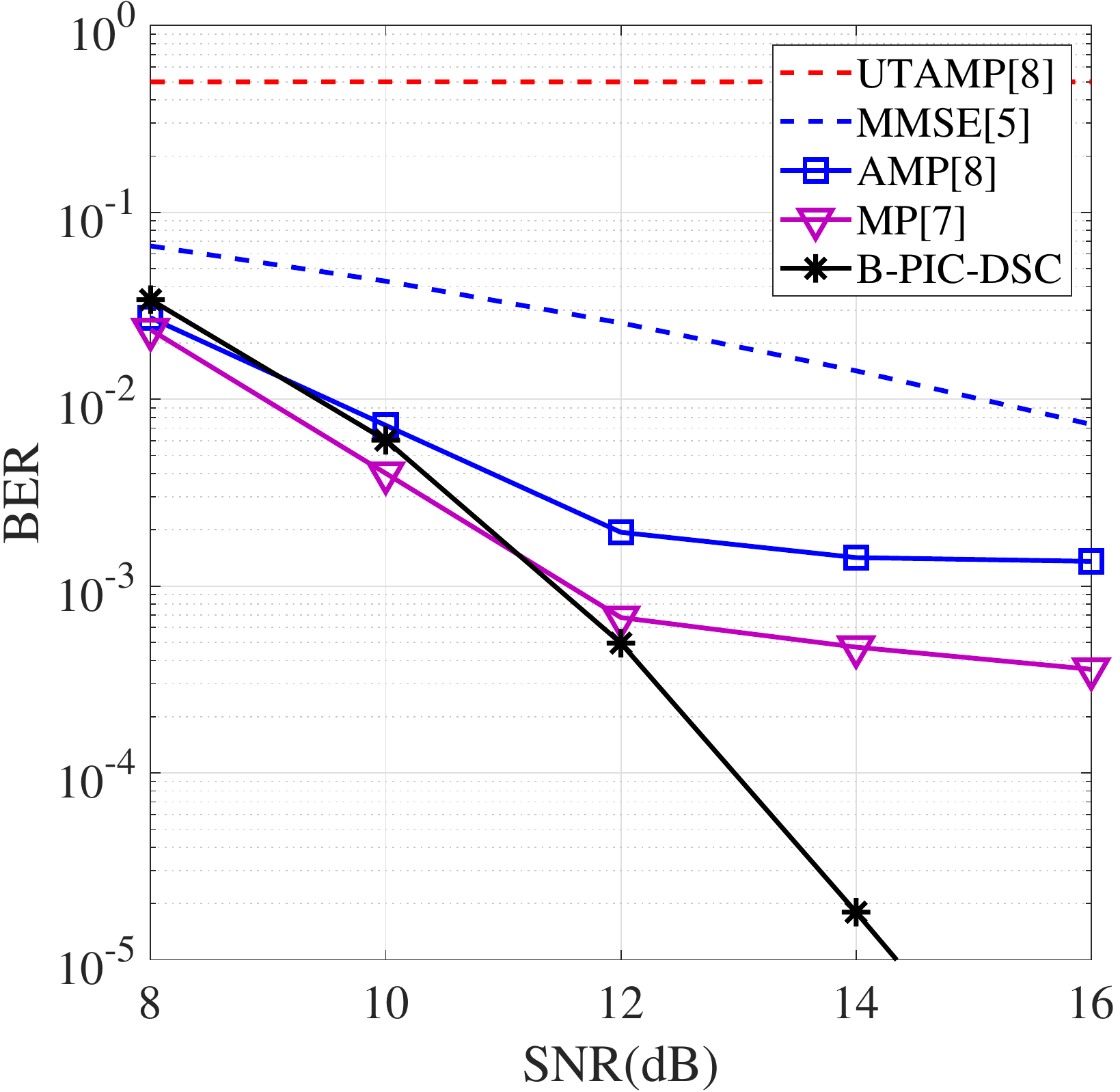}\hfill}
\caption{A large number of mobile reflectors with $k_{max}=3$}
\label{Fig:IB-PIC-DSC-Result-bad-richscattering}
\end{figure}

\section{Conclusion}
We propose the B-PIC-DSC detector for the OTFS systems that is able to achieve a high detection performance, particularly under high ICI  environment, where there is a large number of moving reflectors. Our simulation results show that the B-PIC-DSC OTFS detector outperforms the state-of-the-arts OTFS detectors with slightly higher computational complexity.

\section*{Acknowledgment}
This research was supported by the research training program stipend from The University of Sydney. The work of Branka Vucetic was supported in part by the Australian Research Council Laureate Fellowship grant number FL160100032.

{\renewcommand{\baselinestretch}{1.1}
\begin{footnotesize}
\bibliographystyle{IEEEtran}
\bibliography{IEEEabrv,myBib}
\end{footnotesize}}

\end{document}